\documentclass[12pt]{article}
\usepackage{a4}
\usepackage{array}
\usepackage{fancybox}
\usepackage{hhline}
\usepackage{amsmath,amssymb}
\usepackage[dvips]{graphics}
\usepackage{graphicx}  
\def\re{\hbox{$\mathcal E $}}
\def\rx{\hbox{$\mathcal X $}}
\def\ry{\hbox{$\mathcal Y $}}
\def\rz{\hbox{$\mathcal Z $}}
\def\rf{\hbox{$\mathcal F $}}

\def\ra{\hbox{$\mathcal A $}}
\def\rb{\hbox{$\mathcal B $}}
\def\va{\overrightarrow{A}}
\def\vb{\overrightarrow{B}}
\begin{document}
\title{A fast and precise method to solve the Altarelli-Parisi equations
in $x$ space.}
\author{ C. Pascaud and F. Zomer\\
{\small Laboratoire de l'Acc\'el\'erateur Lin\'eaire, IN2P3-CNRS}\\
{\small et Universit\'e de Paris-Sud, F-91405 Orsay cedex, France.} }
\maketitle

\begin{abstract}
A numerical method to solve
 linear integro-differential equations is presented. This
method has been used to solve the QCD Altarelli-Parisi
 evolution equations within the H1 Collaboration at DESY-Hamburg.
 Mathematical aspects and numerical approximations are described. 
The precision of the method is 
discussed.
\end{abstract}
\vspace{2cm}
{\it 
This article is an extended version of an unpublished note \cite{auto-cite}.

In a recent publication \cite{rat}, P. Ratcliffe proposed   
 a numerical method similar to the one described
  in ref. \cite{auto-cite}. In addition, he
 pointed out the problem of 
 non commutativity of the Next-to-Leading-Logarithmic-Approximation (NLLA)
 Altarelli-Parisi kernels 
 and the fact that we did not account for in
 ref. \cite{auto-cite}. Our present extension of \cite{auto-cite} 
therefore concerns the account for non commutativity.

 After having given
 the proper modification of the numerical method (section 2.1) 
 we explicitly show that non commutativity effects can safely be neglected 
 provided the $Q^2$ evolution is performed, as usual, from points to points on
 a grid (section 2.2). The rest of the paper is untouched, in particular
 the references are not updated.}


\section*{Introduction} 
Inclusive 
Deep Inelastic lepton-hadron Scattering (DIS)cross section measurements offer
 a powerful test of 
perturbative Quantum Chromo Dynamics (pQCD) \cite{ap}. 
 The DIS process $\ell(k)^{\pm}h(P)\rightarrow \ell^{\pm}(k')X$ (here we
 shall only consider the case of charged
leptons $\ell^{\pm}$ in order to simplify the discussion) kinematic is 
described by two Lorentz invariants.
One usually chooses the transferred momentum squared
 $Q^2=-q^2\equiv (k'-k)^2$ and 
 the Bjorken variable $x=Q^2/(2P.q)$. In terms of these two
kinematic variables, the internal dynamics of the struck hadron enters
 the cross section via three structure
functions: $F_1(x,Q^2)$, $F_2(x,Q^2)$ and $F_3(x,Q^2)$. Because
 only $F_2(x,Q^2)$ contributes significantly to
 the cross section for $Q^2 < M_{Z_0}^2c^4$, we shall concentrate
 on this structure function in the present paper.   

Within the framework of pQCD, 
 $F_2(x,Q^2)$ is given by the convolution in $x$ of the well known
 Wilson coefficients \cite{petronzio}
 and of the parton densities inside the hadron (we shall work in the 
$\overline{MS}$ factorisation and renormalisation scheme).
 The densities of partons, consisting of quarks and gluons,
  are computed from the solution of the Altarelli-Parisi (AP)
 equations \cite{ap}. According to \cite{petronzio}, it is useful
 to define the gluon $g$, a non-singlet $q_{\scriptstyle\scriptstyle NS}$
 and a singlet $\Sigma$ quark combination
 densities. They are the solution of the set of AP
 integro-differential equations:
\begin{equation}\label{ns}
\cfrac{\partial q_{\scriptstyle\scriptstyle NS}}{\partial t}
=\int_x^1 \dfrac{dw}{w}P_{\scriptstyle\scriptstyle NS}(w,t)
q_{\scriptstyle\scriptstyle NS}(\tfrac{x}{w},t)
\end{equation}
\begin{equation}\label{gs}
\cfrac{\partial}{\partial t }
\begin{pmatrix}
\Sigma(x,t)\\
g(x,t)
\end{pmatrix}
= \int_x^1 \cfrac{dw}{w}
\begin{pmatrix}
P_{qq}(w,t) & n_f P_{qg}(w,t)\\
P_{gq}(w,t) &  P_{gg}(w,t)\\
\end{pmatrix}
\begin{pmatrix}
\Sigma(x/w,t)&\\
g(x/w,t)&
\end{pmatrix}
\end{equation}
where all kernels $P$ are expanded perturbatively:
\begin{equation}\label{splitting}
P=\alpha_s(t)P^{[1]}(w)+\alpha_s^2(t)P^{[2]}(w)+O(\alpha_s^3)
\end{equation}
Here $t=\log(Q^2/\Lambda^2)$ and $\Lambda\approx 200MeV$ is the pQCD scale parameter. 
  Expressions of the leading and next leading order splitting functions,
$P^{[1]}$ and $P^{[2]}$, can be found in \cite{furman1}.
 The expression of the strong coupling constant $\alpha_s$ from ref.
 \cite{marciano} will be used. 
 However, eq. (\ref{ns}),(\ref{gs}) hold only for $Q^2 \gg \Lambda^2$ where
 the perturbative series (\ref{splitting}) is convergent.
  Some non-perturbative input functions are then required to solve the system.
 In practice, these functions depend on 
unknown parameters (except in the case of \cite{ckm}) which are determined from a fit to
experimental structure function measurements \cite{tung}. Thus, eq. (\ref{ns}),(\ref{gs})
 must be solved many times during the usual
$\chi^2$ minimisation procedure and a fast and precise numerical method is then required.

 Two \footnotemark different numerical methods have been used so far to solve
 eq. (\ref{ns}),(\ref{gs})
(see ref. \cite{chyla} for a review). 
\footnotetext{ \it Analysis of  $F_2(x,Q^2)$
 moments will not be considered here. It is well known since a long
 time \cite{furmanski-cern} that it
 relies too much on
the behaviour of the structure function in the elastic $x\approx 1$ and
 `small-$x$' $x\rightarrow 0$ regions.}    
The first one
uses the fact that the Mellin transform of the system (\ref{ns},\ref{gs})
 leads to a simple set of first order
differential equations in the $x$ complex conjugate moment space. 
The solution is straightforward, after some further necessary
simplification of the evolution kernels, but the price to pay is to
 perform numerically an inverse Mellin transform.
The method proposed in the present
 article belongs to a second kind of 
approach in which the system is solved in the $x$ space:
let us comment with more details the main
 features of three existent approaches of this kind.
\begin{itemize}
\item
It is first natural to use methods based on the Taylor expansion of the
 parton densities in $\log(Q^2)$
  \cite{taylor,virchaux-these,jap1} 
 (called `brutal force' method in ref \cite{chyla,jap1}). But it is well
 known \cite{checo} that such techniques
 lead to rounding errors when the $Q^2$ evolution covers many order of magnitude.
 And most of all,
 reaching a good accuracy is prohibited by a necessary high CPU (see ref. \cite{jap1}
 for a complete 
numerical study). 
 \item
 It was early observed \cite{jacobi} that 
shifted Jacobi polynomials \cite{poly} offer many useful properties to solve AP equations:
 orthogonality for $x\in [0,1]$,
 a weight function $x^\beta(1-x)^\alpha$ describing the asymptotic behaviour 
of parton densities ($\alpha>0$, $\beta>-1$ are real arbitrary numbers)
 To understand advantages and disadvantages
of that method, let us write the expansion of the non singlet densities in terms of
 the shifted Jacobi
polynomials $\Theta^{\alpha,\beta}_n(x)$:
\[
xq_{NS}(x,t)=x^\beta(1-x)^\alpha\sum_{n=0}^{\infty}a_n(t)\times 
\Theta^{\alpha,\beta}_n(x) \hbox{   }.
\]
where $\Theta^{\alpha,\beta}_n(x)=\sum_{j=0}^n c_{j,n}(\alpha,\beta)x^j$. 
Using orthogonality relation one immediately relates the coefficients $a_n(t)$ to
 the non singlet moments.
Although the solution can be explained in a very compact form, 
the sum in the previous equation must
be truncated \cite{jac2}: when taking for example the smallest 
experimental $x=10^{-5}$ accessible value \cite{H1-Zeus}, one sees that increasing
 $n$ leads to 
numerical rounding errors. Note that this method was originally used for analysing
 data at $x>10^{-2}$ and,
as pointed out in \cite{chyla}, `a careful study of numerical stability is required
 to extend
this method at lower $x$ values'. Recently, it was claimed \cite{new-jacobi} that
 this method could
be applied down to $x\approx 10^{-4}$ but no numerical stability studies have been yet
provided. 
\item
In ref. \cite{petronzio-lagrange}, the authors made a series expansion
of both parton densities and kernels. Changing the integration variable $x$ to 
$\log(1/x)$ they show that the best polynomial basis to perform this expansion 
are the Laguerre orthogonal Polynomials \cite{poly}. In particular, the convolution
 product property
of these polynomials leads to an explicate mathematical form of the solution.
 Using the generalised
Laguerre polynomials, they also include the asymptotic behaviour of the parton
 densities as depicted 
in the previous item. The 
restrictions of this method are twofold: 
 the kernels and the parton densities are approximated
  leading to series in the solution which must be truncated.  Furthermore,
 some recent studies of the numerical precision \cite{jap1,jap2} have shown that this
method `may not be satisfactory at small-$x$ and large-$x$' \cite{jap1}.

\end{itemize} 

With the forthcoming high precision $F_2$  measurements by the HERA experiments
\cite{H1-Zeus}, together with the very accurate fixed target experiment
 data\cite{NMC,bcdms,slac},
 precise solutions of AP equations are required in a kinematic plane
 covering five orders of magnitude
in $x$ and $Q^2$. In order to perform QCD analysis in this large kinematic
 domain, we present
a new numerical method to solve the system (\ref{ns},\ref{gs}).
 Using the finite element method, together with a scaling property of the
 convolution integrals 
appearing in eq. (\ref{ns}),(\ref{gs}), we show that it is possible to 
obtain an explicit formula for the solution of AP equations without any infinite series. 
We also show that our formulation leads fortunately to a reduction of the CPU time 
by one order of magnitude with respect to the `brutal' method described in \cite{jap1}.

The rest of this article in organised as follows.
The method is described in sect.~\ref{section1}: 
 notations and formalism
of \cite{deboor} are used. 
Application to the 
AP evolution equations is described in sect.~\ref{section2}.
The numerical precision,
 together with the performances are studied in sect.~\ref{section3}.

\section{Description of the method }\label{section1}

Let us write generically AP integro-differential equations (\ref{ns}),(\ref{gs})
  in the form:
\begin{equation}\label{intdif}
\cfrac{\partial F(x,t)}{\partial t} = \int_{x}^{1}\cfrac{dw}{w} K(\tfrac{x}{w},t)F(w,t),
\quad
K(\tfrac{x}{w},t)=\sum_{m=1}^{2} \alpha_s^m(t)P^{[m]}(\tfrac{x}{w}) \quad ,
\end{equation} 
where $F$ stands for quark or gluon 
densities in the proton and therefore belongs to a class of smooth functions.
 In addition, one has $x \in [0,1]$, $t \in [0,\infty]$ and $F(1,t)=0$ for all $t$.

We start to solve eq. (\ref{intdif}) by defining an ordered sequence of 
 points \footnotemark$|x>=\{x_i\}_1^n$ and the corresponding sequence 
$|F>=\{F_i\}_1^n$ with $F_i=F(x_i,t)$ . It is then possible to define a 
piecewise polynomial function $\mathfrak F_r(x)$ of order $r$ which interpolates
$|F>$~in $x$ at a given value of~$t$:~$\mathfrak F_r(x_i)=F_i$. \\
 $\mathfrak F ^r$ is described by a set of polynomial
functions $|\mathfrak P>=\{\mathfrak P^r_i\}_1^{n-1}$ of order $r$: 
 $$\mathfrak F_r(x)=\mathfrak P^r_i(x) \hbox{    if    } x_i\le x \le x_{i+1}.$$ 
 $\mathfrak F_r$ together with $|x>$ define a linear space 
$\mathbb P_{r,x}$. If one supplies a sequence of continuity conditions
$|\nu>=\{\nu_i\}_2^{n-1}$, the defined space $\mathbb P_{r,x,\nu}$ is still
linear and is a subspace of $\mathbb P_{r,x}$. Because of the linearity of
$\mathbb P_{r,x}$, one can define a basis of continuous functions  
$|\phi>=\{\phi_i(x)\}_1^{n}$ such that $\phi_i(x_j)=\delta_{ij}$ 
($\delta_{ij}$ stands for the kroeneker symbol). In this
way, one can write 
\begin{equation}\label{expen}
\mathfrak F_r(x)=\sum_{i=1}^{n}F_i\phi_i(x),
\end{equation}
\footnotetext{\it We use the following convention: 
 $\{x_i\}_1^n \equiv(x_1,...,x_n)$.}

with $|\phi_i> \in \mathbb P_{r,x,\nu}$. Equation (\ref{expen}) is valid if and only
if 
\begin{equation}\label{dim}
dim\hbox{  }\mathbb P_{r,x,\nu} \equiv r\times n-\sum_{j=2}^{n-1}\nu_j=n \quad ,
\end{equation}
and is thus only satisfied for a restrictive class
of sequences $|\nu>$ . Functions $\mathfrak F_r$ which are 
constructed in such a way are the original spline functions\cite{deboor}. 

We turn now to the solution of (\ref{intdif}) by making two assumptions:
$$\mathfrak F_r(x)=F(x) \hbox{ and }
 \int_{x}^{1}\cfrac{dw}{w} K(x/w)\phi_i(w) \in \mathbb P_{r,x,\nu}.$$
 Depending on the choice of $|x>$ and $r$,
 theorems concerning the precision of these approximations are given in 
\cite{deboor}. This topic will be covered in sect.~\ref{section3}
 in the case of AP equations.

Using (\ref{dim}), eq. (\ref{intdif})
 takes a matrix form
\begin{equation}\label{intdifmat}
\cfrac{\partial}{\partial t}|F> = 
\sum_{m=1}^2 \alpha_s^m(t)M_m|F>,
\end{equation}
with 
\begin{equation}\label{matint}
<i|M_m|j>=\int_{x_i}^{1}\cfrac{dw}{w} P^{[m]}(x_i/w)\phi_j(w),
\end{equation}
The solution of (\ref{intdifmat}) can therefore be written formally
\begin{equation}\label{mainsol}
|F>= \exp \biggl( \sum_{m=1}^2\int_{t_0}^{t}dt'\alpha_s^m(t') M_m\biggr)|F_0>,
\end{equation}
where $|F>=|F_0>$ at $t=t_0$ ($t_0>0$ can be chosen arbitrary).

Next we choose the sequence $|x>$ and the basis functions $|\phi>$
according to the condition (\ref{expen}). We also focus on the numerical
computation aspects, namely the CPU time and the precision. 
 
We first remark that if 
\[
x_{i+1}=\lambda x_i, \hbox{  with  }\lambda >1 
\hbox{  and  }
\begin{cases}
 \phi_{i}(x)& =\phi_{i+1}(\lambda x), \\
 \phi_{i}(x)& =0 \hbox{ for } x\ge x_{i+1}, \\
\end{cases}
\]
matrices $M_m$ are triangular and $<i+1|M_m|j+1>=<i|M_m|j>$.
This very interesting scaling property relies on the fact that the integrations
are convolution products running from
$x$ to 1; it remains true even in the 
presence of a `$+$' (QCD) prescription appearing in the AP kernels\cite{ap}.

As we are free to choose the order $r$ of the piecewise polynomials,
 we take the simplest case, {\it i.e}
 linear interpolation ($r=2$). It is known to be 
 nearly optimal for large $n$ values. The corresponding basis is unique and
 consists on the Lagrange (or sometime called `hat') functions:
\begin{equation}\label{phi}
\phi_i(x)=
\begin{cases}
(x-x_{i-1})/(x_i-x_{i-1}) & \hbox{,   }x_{i-1}\le x \le x_i \\
(x_{i+1}-x)/(x_{i+1}-x_i) & \hbox{,   }x_i\le x \le x_{i+1} \\
0                         & \hbox{,   otherwise}.
\end{cases}
\end{equation}
 Note that integrals of eq. (\ref{matint})
are simple to compute using definition (\ref{phi}); cancellations of singularities 
 contained in the AP kernels can even be computed analytically.
 
If one wishes to design a computer program it is necessary to define completely
the points where the function $F$ has to be calculated. This is done by
 defining a sequence in the variable $t$: $|t>=\{t_k\}_1^{n_t}$.
The solution may then be propagated from the beginning of the grid to its end
by using a recursive expression in analogy with eq.~(\ref{mainsol}):
\begin{equation}\label{recur}
|F_{k}> = \exp{A_k |F_{k-1}>} \hbox{  with  } A_k =
\sum_{m=1}^{2}\int_{t_{k-1}}^{t_{k}}dt'\alpha_s^m(t') M_m \hbox{  ,  }k=1,\ldots,n_t\quad.
\end{equation}
In the case of a fitting procedure, if the kernel $K(w,t)$ is not changed 
({\it i.e} $\alpha_s$ fixed at a certain scale $t$)
it is sufficient to compute the matrices $\exp(A_k)$
only once. Furthermore, the type of integrals appearing in eq. \ref{matint} are computed
at the initialisation step because they depend on the choice of $|x>$ only.  
They can be computed with a high numerical precision.
If only $|F_0>$ is changed one must only redo the matrix
multiplications implied by eq. (\ref{recur}): this is a consequence of the
linearity of (\ref{intdif}).
 
 To efficiently exponentiate matrices $A_k$, it is powerful to introduce
the `band' matrices $B$ defined by:
$$<i|B_l|j>=\delta_{i+l,j}\quad ,\quad l\ge 0\quad.$$
It is easy to show that $B$ matrices fulfil the multiplication rule 
$$B_iB_j=B_jB_i=B_{i+j}.$$
 These matrices form a
basis for the kernels matrices $A_k$ and we can
write 
$$A_k=\sum_{l=0}^{n-1}{ }^ka_lB_l \hbox{   with   } { }^ka_l=
\sum_{m=1}^2\int_{t_{k-1}}^{t_{k}}dt'\alpha_s^m(t') <l|M_m|1>.$$
The product of two matrices such as 
$A_k$ will be a matrix of the same type, {\it i.e}
 a linear combination of the $B$'s.
The use of band matrices leads to a number of operations of 
order $n^2$ whereas ordinary matrices would give a CPU time increasing as $n^3$.

Let us now isolate the diagonal term in $A_k$ and rewrite:
$$A_k=\sum_{i=0}^{n-1}{ }^ka_iB_i={ }^ka_0B_0+A_k'.$$
Commutativity allows us to  split the exponential in two
parts:
$\exp(A_k)= \exp({ }^ka_0)\exp(A_k').$
 In this expression the second exponential can be expanded as
a series which will consists in a sum of products of $B$ matrices. As
$B_0$ is not an element of those products and because of the multiplication
rule  the index of the
$B$'s will increase along the sum to reach its maximum $n-1$. So
there will be a limited number of terms in the series and one can
write 
\begin{equation}\label{expeq}
\exp(A_k)=\exp({ }^ka_0)\sum_{l=0}^{n-1}\cfrac{(A_k')^l}{l!} 
=\sum_{j=0}^{n-1} { }^kT_j B_j.
\end{equation}

If we turn back to eq. (\ref{intdifmat}), we can finally write the solution
in the following form:
\begin{equation}\label{solution}
F(x_i,t_{k+1})=\sum_{j=0}^{n-i}{ }^kT_j F(x_{j+i},t_k).
\end{equation}
This equation shows explicitly that $F(x_i,Q^2)$ depends only on
the values of the function $F$ for $x>x_i$. 

\section{Application to the Altarelli-Parisi equations}\label{section2}

It is clear that the non-singlet equation (\ref{solution})
 is a solution of eq. (\ref{ns}). 
 In the case of singlet and gluon coupled differential
equations, we may still retain formally eq. (\ref{mainsol})
but $A$  is now made of four sub matrices.
Let us rewrite eq. (\ref{mainsol})  with slightly different notations:
\begin{equation}\label{vecsol}
|\rf_{k+1}>= e^{\ra k} |\rf_k> \hbox{     with        }
\ra = \begin{pmatrix}
A_{qq} & A_{qg}\\
A_{gq} & A_{gg}\\
\end{pmatrix},
\end{equation}
where the \ra  matrix can longer be expended linearly on the 
 basis of band matrices,
 unlike $A_{qq}$, $A_{qg}$, $A_{gq}$ and $A_{gg}$.
We can also consider that the vectorial space involved is a direct
product of a n-space and of a 2-space and write (see Appendix A):
\begin{equation}\nonumber
\ra= A_0 \re +\va  = A_0 \re + A_x \rx+ A_y \ry+ A_z \rz \quad ,
\end{equation}
where 
$$ \re=
\begin{pmatrix}
1 & 0 \\
0 & 1
\end{pmatrix}\hbox{ },\quad
\rx=
\begin{pmatrix}
0 & 1 \\
1 & 0 
\end{pmatrix}\hbox{ },\quad
\ry=
\begin{pmatrix}
0 & 1 \\
-1 & 0 
\end{pmatrix}\hbox{ },\quad
 \rz=
\begin{pmatrix}
1 & 0 \\
0 & -1
\end{pmatrix}\hbox{ }.
$$
The following rules hold:
\begin{align}\label{multrule}
 \re &= \re^2 = \rx^2 = -\ry^2 = \rz^2\,,\\
\rx \ry &=\rz \,\,\hbox{ + permutations}\,,\\
\end{align}
$\re$ commutes with $\rx$, $\ry$, $\rz$ and those anti commute
between themselves.
Therefore, one can split into pieces the exponential 
\begin{equation}\nonumber
e^{\ra } = e^{A_0 } e^{ \va }\hbox{  }.
\end{equation}
The second exponential in the r.h.s of the above equation 
is computed by series expansion. Then, one 
needs to calculate  powers of  $ \va $.
Making use of eq (\ref{multrule}) one gets (see Appendix A):
\begin{equation}\nonumber
(\va)^2 =(A_x \rx+ A_y \ry+ A_z \rz)^2 = 
(A_x^2 -A_y^2 + A_z^2)\re = A_e^2 \re.
\end{equation}
This shows that we can split the series in its even and
 odd parts and sum them independently. The result is the following:
\begin{equation}\label{relation}
e^{\ra } = e^{A_0 } \left(
\cfrac{{e^{A_e }+e^{-A_e}}}{2}\re
+ \cfrac{{e^{A_e}-e^{-A_e}}}{2}A_e^{-1}\va
\right).
\end{equation}
 $A_e^2$, its square
root $A_e$ and the inverse are easily computed \footnotemark
 using the $B$'s multiplication rules.
\footnotetext{\it Depending on the sign of  $A_e^2$ 's
diagonal, $A_e$ will be real or pure imaginary; it turns out for AP equations
that we are in the first case.}

\subsection{Non commutativity correction}

Eq. \ref{vecsol} is a solution of the AP equations if the following relation
holds:
$$\frac{\partial}{\partial t} e^{\ra } = \frac{\partial}{\partial t}
\ra \times  e^{\ra } = \ra '  \times  e^{\ra }\,.$$
Looking at the series expansion of the exponential and its derivative one
can see that it is the case only at the leading order in $\alpha_s$ where
$\frac{\partial}{\partial t} \ra$ and $\ra$ commute.

In the general case, i.e. including higher orders is $\alpha_s$,
 the same form for the solution may be kept by defining an
operator $\rb$ such that
$$ \biggl\{
\frac{\partial |\rf>}{\partial t}=\ra  |\rf>\,\,{\rm and}\,\,|\rf>= e^{\rb } |\rf> 
\biggr\}
\Rightarrow
\frac{\partial}{\partial t} e^{\rb } =
\ra ' \times  e^{\rb }$$
Simple but tedious non commutative algebra (see Appendix A) gives:
\begin{align}
B'_0 -A'_0 &= 0\nonumber\\
\vb'- \va' &= H \va' + \overrightarrow{\va' \vb} 
-  \frac{H}{B^2}(\vb \va')_0 \vb\,.\nonumber
\end{align}
where $H$ is given by eq. \ref{hh}.
As those correction terms are always small (see below)
integration can be made by a simple trapezoidal formula onto a $Q^2$ grid
and the system can be
solved iteratively using $\ra'$, $\ra$ as initial value for $\rb'$, $\rb$.

\subsection{Effect of the non commutativity correction}

To check the validity of this approach the singlet density have been evolved 
in the NLLA according to four different assumptions:
\begin{enumerate}
\item
the solution is propagated from one point of a $Q^2$ grid to the next one as
usually done and the non commutativity correction is not applied.
\item
the solution is propagated from one point of the $Q^2$ grid to the next one and
the non commutativity correction is applied. This method should be the best one
and is used has the basis of the comparison.
\item
the solution is propagated from the first point of the $Q^2$ grid to the actual
one and the non commutativity correction is not applied. No $Q^2$ grid is considered.
\item
the solution is propagated from  the first point of the $Q^2$ grid  to the actual
one and the non commutativity correction is applied. No $Q^2$ grid is considered.
\end{enumerate}
To compare the different results we perform the ratio of the 
 singlet densities $\Sigma_2(x,Q^2)/\Sigma_1(x,Q^2)$,
 $\Sigma_3(x,Q^2)/\Sigma_1(x,Q^2)$ and $\Sigma_4(x,Q^2)/\Sigma_1(x,Q^2)$ where the 
 subscript refers to the above assumptions. As for the gluon density, the effect is
 found to be an order of magnitude smaller than for the singlet case.
   
\begin{figure}
\begin{center}
\includegraphics[height=12cm,width=12cm]{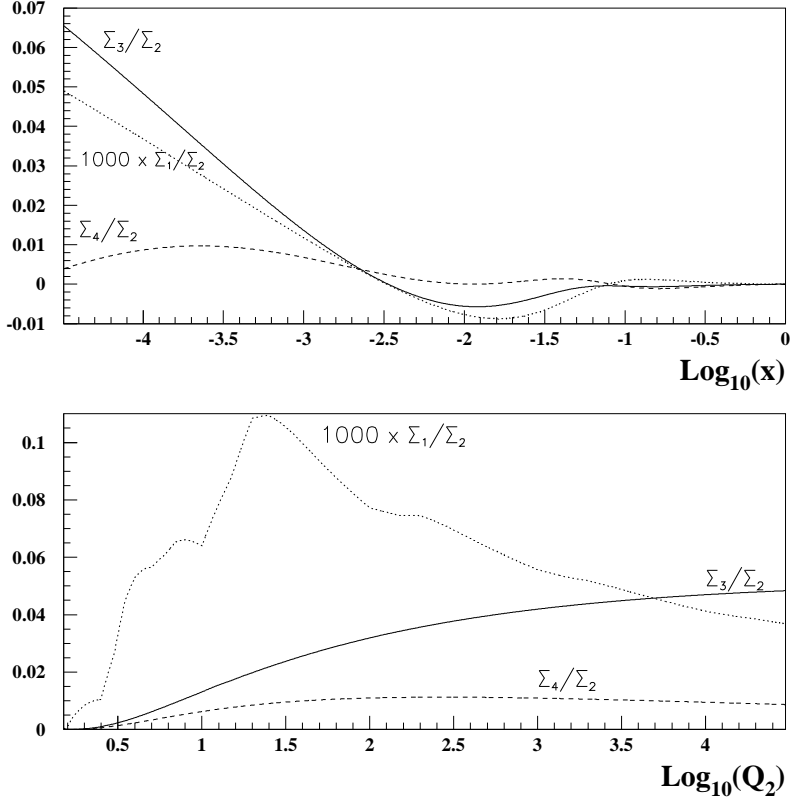}
\end{center}
\caption{\it  Non commutative corrections, see text.}
\label{commute}
\end{figure}

Figure \ref{commute} shows the singlet ratios as function of $x$ at
 $Q^2=30000$~GeV$^2$ and as
function of $Q^2$ at $x=0.0001$ ($\Sigma_2(x,Q^2)/\Sigma_1(x,Q^2)$ is scaled by a
 factor 1000 in this figure). The starting scale of the $Q^2$ evolution is 1 GeV$^2$.
 The parametrisation of the gluon and singlet densities at this scale is taken from
 a recent global fit \cite{nousb}. 

As expected method 1 and 2 are very close, the bias being less than $10^{-4}$
with an irregular shape showing that it comes entirely from rounding errors.
 For method 3 the bias is noticible at small $x$ and 
large $Q^2$ where it goes up to 6.5 \%. When the 
non commutativity correction is applied (method 4 ) the bias is under
control and less than 1 \% \footnotemark .
\footnotetext{ \it This remaining bias is independent of the $x$ and $Q^2$
grid sizes and seems due to computer accuracy (the number of convolution
products needed being realitvely high)}

{\bf This study shows that, if one propagates the Altarelli-Parisi
 solutions from point to point
on a $Q^2$ grid, the non commutativity bias is completely negligible and that
it is not even necessary to apply the correction.}

\section{Precision and performances}\label{section3}

The precision of the method proposed in this paper can be easily
estimated by changing the number $n$ of breakpoints in $|x>$.
The convergence of the evolved parton densities leads to an estimation of the precision. 
Input functions and
$\Lambda$  ($\Lambda^4=225MeV$) value of \cite{H1_PAPER}
are used:
\begin{align*}
xg(x)&= 1.86x^{-0.22}(1-x)^{7.12} \\
x\Sigma(x)&= 1.15x^{-0.11}(1-x)^{3.10}(1+3.12x)\\
xq_{NS}(x)&= 1.14x^{0.65}(1-x)^{4.66}(1+8.68x)
\end{align*}
 at $Q^2_0=4\hbox{ }GeV^2$. Here, 
 $\Sigma=u+\bar{u}+d+\bar{d}+s+\bar{s}$ and
$q_{NS}=u+\bar{u}-\Sigma/3$ is the chosen non-singlet quark density
(for more details concerning the fit we refer to \cite{H1_PAPER}).

We define the two 
bounded sequences, $|x>$ such that $x_i \in [x_{min}=10^{-5},1], i=1, \ldots,n$
, and $|t>$ such that $t_j \in [0,10],j=1, \ldots,n_t$. 
 We start from $n=100$ and increase up to $n=900$ by steps of 100.
  $n_t$ is set to $100$ for all $n$ and this does not introduce numerical
 uncertainties since the expression of eq. (\ref{expeq}) is exact.  
Evolutions of eq. (\ref{ns}),(\ref{gs}) are performed 
up to $Q^2=10^4GeV^2$. Fig. (\ref{lafigure}a) shows
 the ratio $R_{q_{\scriptstyle\scriptstyle NS}}(x,n)$ defined by:
\begin{equation}\label{r}
R_{q_{\scriptstyle\scriptstyle NS}}(x,n) = 
\cfrac{{q_{\scriptstyle\scriptstyle NS}}(x)|_{n,n_t}}
{{q_{\scriptstyle\scriptstyle NS}}(x)|_{n=900,n_t}}
\hbox{  ,   for   }n=800,700,600,500,400,300,200,100
\end{equation}
for $Q^2=10^4GeV^2$. Note that the calculations done with large values
 of $n$ are stable because of a choice of
a first order spline interpolations (see section \ref{section1}). With
 higher order splines, the choice of the
breakpoint set $|x>$ is very important and solutions may be unstable \cite{deboor}.
 Figs.(\ref{lafigure}b,c) show
$R_\Sigma(x,n)$ and $R_g(x,n)$ respectively. These plots illustrate the
convergence of the method. 
The wiggles appearing in these plots are generated by the Lagrange functions
 which are also
used to interpolate between the breakpoints of the sequence $|x>$.
From these plots, one sees that for $n=300$, the precision
is of the order of $0.5\%$ in the `low-$x$' region. This is accurate enough 
comparing to 
 measurement uncertainties \cite{H1-Zeus},\cite{NMC}.
 However, at `high $x$', huge differences appear. Although these instabilities
 of the calculations do not
modify the results in the `low-$x$' region, one cannot use this method to compute
the structure functions at `low-$x$' and `high $x$' simultaneously with a good
 precision.
 A straightforward modification of the method is to define a second net in $x$.
 This new net starts from
 $x'_{min} = x_{min}^{1/n'}$. Then 
$n'$ breakpoints are set equidistantly in $\log(x)$ between two adjacent points
 of the first net up to $x=1$. 
For the calculations,
the second net is first used to solve the AP equations. The extension from 
$x'_{min}$ down to $x_{min}$
is then made by solving AP equations using the first net. For example, taking
 $x_{min}=10^{-5}$
and $n'=5$, the precision for $x\ge x'_{min}=10^{-1}$ will be equivalent to the
 use of one net
made of $n\times n'$ breakpoints. Note that the CPU time is only multiplied by
 two instead of $(n')^2$. 
 To show the resulting precision the fit described above is performed with
 $n=700$ and $n'=1,4,8$ ($x'_{min}=10^{-5},5.6\times 10^{-2},2.4\times 10^{-1}$).
 Then the ratio  
\begin{equation}\label{rprime}
R'_{q_{\scriptstyle\scriptstyle NS}}(x,n',n) = 
\cfrac{{q_{\scriptstyle NS}}(x)|_{n',n,n_t}}{
{q_{\scriptstyle NS}}(x)|_{n'=8,n_{ref}=700,n_t}} 
\end{equation}
is computed. For $n=700$ and $n'=8$, one can reasonably assume that the precision
 is nearly optimum at $x>2.4\times 10^{-1}$.
 We shall take the parton densities, computed with this conditions, as the reference
 `high-$x$' sets. 
 Fig.(\ref{lafigure_2}) show $R'_{q_{\scriptstyle\scriptstyle NS}}$, $R'_{\Sigma}$
 and $R'_g$ for $n=700$:  
 the gain in precision at `high-$x$' is very significant when going from
 $n'=1$ to $n'=4$. 
 On Fig.(\ref{lafigure_2}a,b,c), the ratio $R'$ for $n=250$ and $n'=1,4,8$
 is also shown. 
As pointed out above, $n=250$ leads to 
structure function computations precise enough at `small-x'. One can then
 observe that setting $n'=8$ leads to
great improvement of these computation at `high-$x$'. We conclude that
 for $n=250, n'>4$, it is possible to
perform a very precise calculation of the structure functions within five
 orders of magnitude in $x$ and we point out
that any global pQCD structure function analysis should pay much attention
 to the numerical
precision (as it was already mentioned in \cite{chyla}).

One can design a very fast (CPU) and efficient procedure to determine the
 input functions from a $\chi^2$ minimisation: 
first the minimum is approached rapidly
by setting $n=100$, then the `true' minimum is reached by increasing $n$ 
in steps of $100$ until $n=250$. If `high $x$' measurements enter the fit,
 a second net must be considered at the end of
this procedure.

 According to the AP equations, the momentum sum-rule 
$$
S(t)=\int_{0}^{1} x \biggl(
\Sigma(x,t)+g(x,t) \biggr)dx,  
$$
should stay constant during the evolution in $t$.
 This is a sensitive test of the computational precision.
 With the parametrisations chosen
for the input functions \cite{H1_PAPER}, $S(Q_0^2)$ is computable analytically
and is equal to a sum of Beta functions. $S(Q_0^2)$ is then set to 1
and $S(Q^2)$ is computed numerically with a linear extrapolation in the interval
$x\in [0,10^{-5}]$. This extrapolation introduces a small bias in the 
calculation of $S(Q^2)$.  
Fig. \ref{sumrule} shows $S(Q^2)$ as a function of $Q^2$ for the nine
 sequences of breakpoints of eq. (\ref{r})
 (here one $x$ net is considered). In any case, the deviation from
1 for an evolution in $t$ which covers four orders of magnitude is of the 
order of $2/10000$. 

\section*{Conclusion}

 We have presented a numerical method to solve the Altarelli-Parisi
 equations in $x$ space.
 Unlike conventional methods based on Taylor expansion, the $Q^2$
 evolution is computed 
`exactly'. All convolution products involving
 Altarelli-Parisi Kernels are computed only once with high accuracy.
 Using the Lagrange functions to construct a basis and taking advantage
 of scaling properties
 of AP equations, the number of operations 
 increases with $n^2$ instead of $n^3$ like for other simple basis functions. 
In some senses, our method is equivalent to the polynomial approaches and
 specially the one of 
ref. \cite{petronzio-lagrange}. But, the $x$ space discretization allows
 the use of first order polynomials
to avoid precision computational problems and the freedom in the choice
 of the breakpoint sequence leads
to a formal solution involving only finite series.  
The only numerical instability comes from the number of breakpoints
in the $|x>$ sequence. The resulting numerical precision is found to be very
good when a large number of breakpoints are considered or when two $|x>$ sequences
are used. 
\appendix
\section{Appendix}

\subsection{Notations}
We shall adopt ad hoc, non standard, notations.
 All the components $A_i$ and $B_i$ with $i=0,x,y,z$ can be expended linearly on
 the basis of band matrices.
\begin{itemize}
\item
$[A_i,A_j]=0 \,\,\forall i,j$
\item
$\ra= A_0 \re +\va  = A_0 \re + A_x \rx+ A_y \ry+ A_z \rz $; 
\item
Explicitly:
\[
A_x \rx=
\begin{pmatrix}
0&\dots&0&A_{0,x}&\dots&A_{n,x}\\
\vdots&&\vdots&\vdots&\ddots&\vdots\\
0&\dots&0&0&\dots&A_{0,x}\\
A_{0,x}&\dots&A_{n,x}&0&\dots&0\\
\vdots&\ddots&\vdots&\vdots&&\vdots\\
0&\dots&A_{0,x}&0&\dots&0
\end{pmatrix}
\]
where $A_{i,x}$ are the components of the matrix $A_x$
 (there is only $n$ independent values)
 and $n+1$ is the number of discrete points in the $x$ space.
\item
$\va  =  A_x \rx+ A_y \ry+ A_z \rz $ is an element of the three
dimensional vectorial space generated by $\rx$, $\ry$,$\rz$.
\item
$[\ra,\rb]=[\va,\vb]$.
\item
$(\va \vb)_0=(A_xB_x-A_yB_y+A_zBz)\re$ is the $\re$ component of the
 product of $\va$ and $\vb$.
\item
$\overrightarrow{\va \vb}$ is the vector 
component of the product of $\va$ and $\vb$.  
\item
$ A = \sqrt{(\va \va)_0}\equiv\sqrt{\va \va}=A_e\re$ with
 $A_e^2=A_x^2-A_y^2+A_z^2$.
\item
$\va \wedge \va = \begin{pmatrix}
A_x^2 & -A_x A_y & A_x A_z  \\
A_x A_y & -A_y^2 & A_y A_z \\
 A_x A_z & -A_x A_y & A_z^2 \\
\end{pmatrix}
$ is a matrix in the vectorial space.
\end{itemize}

\subsection{Theorems}

It is easy to demonstrate the following.
\begin{enumerate}
\item
$(\va \vb)_0 =(\vb \va)_0$
\item
$\overrightarrow{\va \vb}= -\overrightarrow{\vb \va}$
\item
$\overrightarrow{\va \va}= 0$
\item
($\overrightarrow{\va \vb}\vb)_0=0$
\item
$(\vb \va)_0 \va = (\va \wedge \va) \vb $
\item
$(\va \wedge \va)^2 = A^2(\va \wedge \va)$
\item
$(1+l \va \wedge \va )^{-1} = 1 - l  \va \wedge \va (1+lA^2)^{-1}$
\end{enumerate}

\subsection{Solution of $ \frac{\partial}{\partial t} e^{\rb } =\ra ' \times  e^{\rb }$}

Using  
\[
 \frac{\partial}{\partial t} e^{\rb } =\ra ' \times  e^{\rb }
\] 
and the commutation relations of the $A_i$ band matrices on gets
\[
(e^{\vb})' =(e^{\rb-B_0})'=(\ra '-B_0')e^{\vb}\,.
\]
We will now assume that $B_0' = A_0'$ and show later its justification.
 As a consequence this equation becomes:
\begin{equation}\label{a0}
(e^{\vb})' = \va ' e^{\vb}
\end{equation}
In order to use eq. \ref{relation}, let us define:
\begin{align}\label{hh}
G &=\frac{{e^{B}+e^{-B}}}{2}\,,\nonumber\\
F &=\frac{{e^{B}-e^{-B}}}{2}B^{-1}\,,\nonumber \\
H &=GF^{-1} -1\,,
\end{align}
so that $G' = FBB' $ and $F' = FH B'B^{-1}$.
 Equation \ref{a0} becomes:
$$FBB' + FH B'B^{-1} \vb + F \vb' = \va' (G+F\vb)$$
or equivalently
\begin{equation}\label{a}
BB' +H B'B^{-1} \vb + \vb' = \va'(1+H +\vb)\,.
\end{equation}
The 0 component reads
\begin{equation}\label{bbp}
BB' = (\vb \va')_0\,,
\end{equation}
and the vector component 
\begin{equation}\label{aa}
H BB'B^{-2}\vb +\vb' = (1+H)\va' + \overrightarrow{\va' \vb}\,.
\end{equation}
Multiplying the two sides of this equation by $\vb$
 and taking the 0 component one gets:
\begin{equation}\label{bbb}
(1+H)BB' =(1+H) (\va' \vb)_0 
+(\overrightarrow{\va' \vb}\vb)_0\,,
\end{equation}
using 
$$BB'=\frac{1}{2}(B^2)'=\frac{1}{2}(\vb\vb)_0'=(\vb\vb')_0\,.$$
The last term of eq. \ref{bbb}
 being nul (theorem 4) we find the result of eq. \ref{bbp}
 (i.e. the 0 component). Hence, among the four equations only three
 are independent: having fixed $B'_0 = A'_0$ we obtain
 a system of three equations with three unknowns (the components of $\vb'$). 
 Therefore, in virtue of the unicity of the solution of a
 first order differential
 equation, $B'_0 = A'_0$ is justified a posteriori.

Equation \ref{aa} may be rewritten
\begin{equation}
\biggl(1+ H B^{-2}\vb \wedge \vb\biggr)\vb'
 = (1+H)\va' + \overrightarrow{\va' \vb}
\end{equation}
and solved with theorem 7
\[\vb' =\biggl(1 - HB^{-2} \vb \wedge \vb(1+H)^{-1}\biggr)\biggl( (1+H) \va'
+  \overrightarrow{\va' \vb} \biggr) \,.\]
A further simplification, using theorem 4, leads to the final expression:
\begin{equation}
\vb'- \va' = H \va' + \overrightarrow{\va' \vb} 
-  HB^{-2}(\vb \va')_0 \vb\,.
\end{equation}

\section*{Acknowledgement}

We thank V. Barone for useful comments and helpful discussions.

\begin{figure}
\includegraphics[height=15cm,width=15cm]{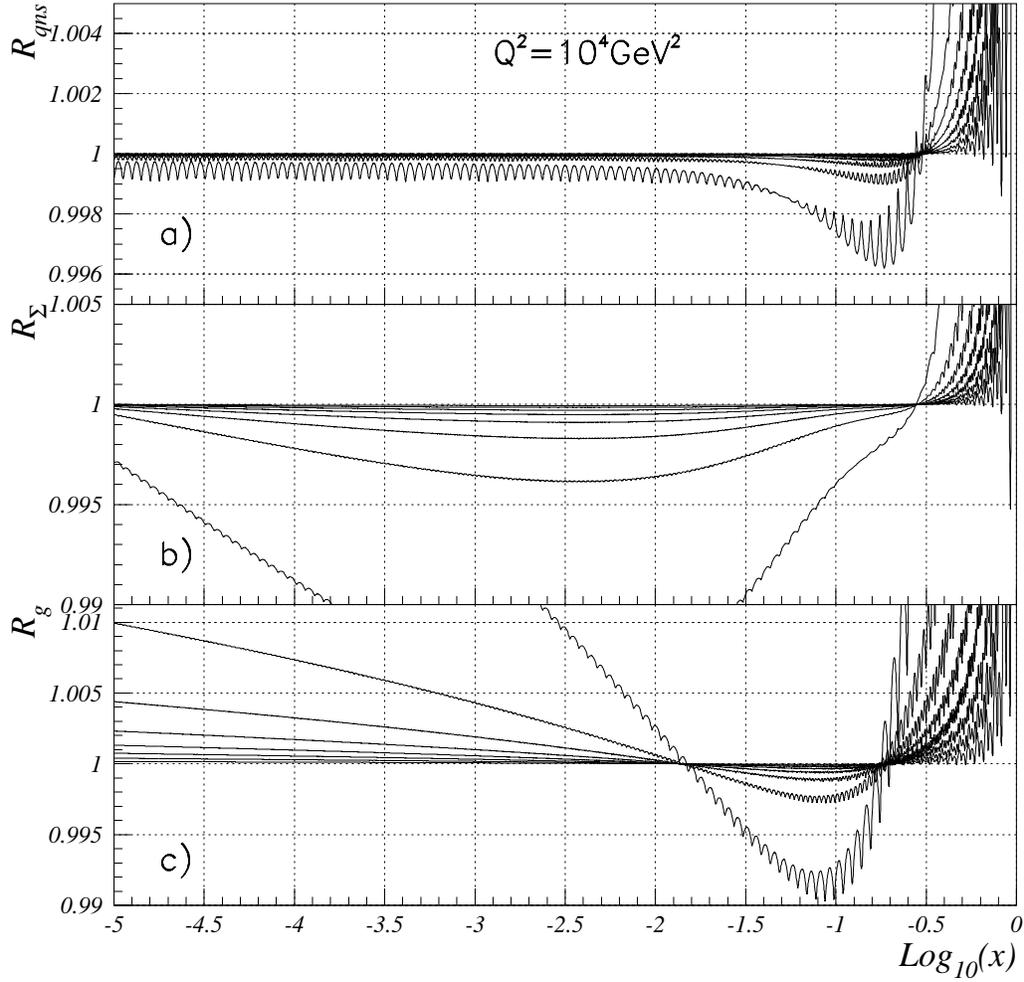}
\caption{\it a)  $R_{q_{\scriptstyle NS}}(x,n)$;
 b) $R_\Sigma(x,n)$; c) $R_g(x,n)$. See eq. (\ref{r}) for definitions.
 The less accurate curve (far from the line $R=1$) corresponds to $n=100$. The
next one to $n=200$ etc... }
\label{lafigure}
\end{figure}

\begin{figure}
\includegraphics[height=15cm,width=15cm]{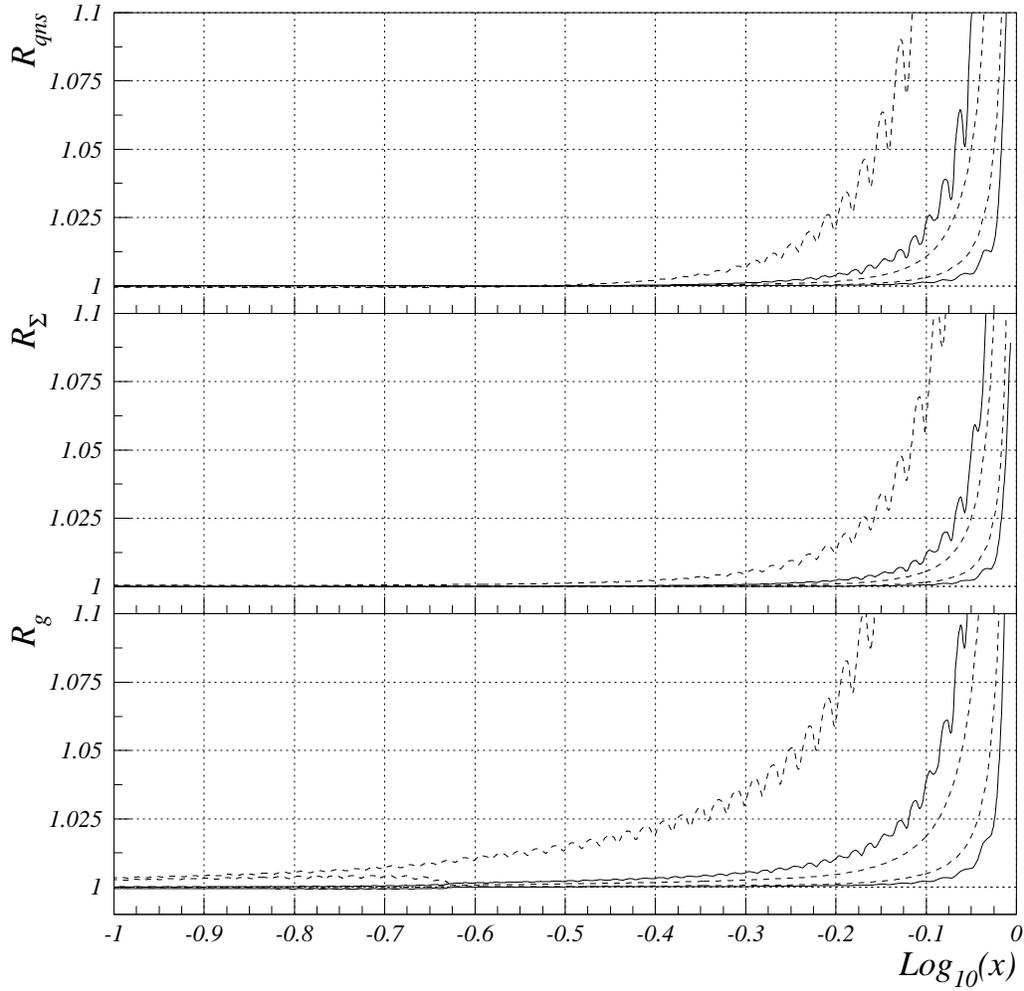}
\caption{\it a)  $R'_{q_{\scriptstyle NS}}(x,n,n')$;
 b) $R'_\Sigma(x,n,n')$; c) $R'_g(x,n,n')$. See eq. (\ref{rprime}) for definitions.
The full lines correspond to $n=700$ and the dashed lines correspond to $n=250$.  
 The less accurate curve (far from the line $R=1$) corresponds to $n'=1$. The
next one to $n'=4$. The last one (defined only in the case $n=250$) to $n'=8$. }
\label{lafigure_2}
\end{figure}

\begin{figure}
\includegraphics[height=15cm,width=15cm]{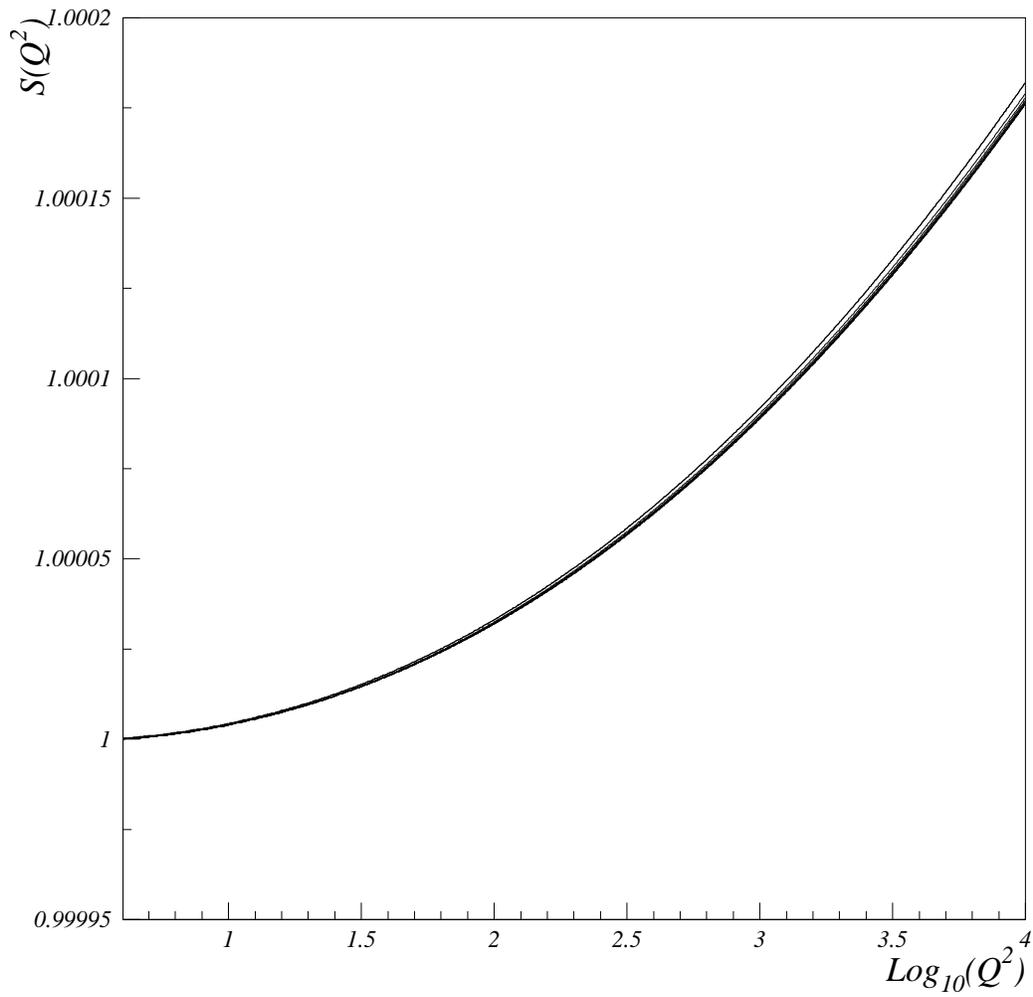}
\caption{\it Momentum sum-rule as function of $Q^2$. The curves described
in the text superpose almost exactly.}
\label{sumrule}
\end{figure}

\end{document}